# Disentangling spin, anomalous and planar Hall effects in ferromagnetic/heavy metal nanostructures

Inge Groen[1], Van Tuong Pham[1,2], Naëmi Leo[1], Alain Marty[2], Luis E. Hueso[1,3], Fèlix Casanova[1,3,*]

[1] CIC nanoGUNE, 20018 Donostia-San Sebastian, Basque Country, Spain
[2] University Grenoble Alpes, CEA, CNRS, G-INP, Spintec, F-38054 Grenoble, France
[3] IKERBASQUE, Basque Foundation for Science, 48013 Bilbao, Basque Country, Spain

[*]Email: f.casanova@nanogune.eu

## Abstract

Ferromagnetic (FM)/heavy metal (HM) nanostructures can be used for the magnetic state readout in the proposed magneto-electric spin-orbit logic by locally injecting a spin-polarized current and measure the spin-to-charge conversion via the spin Hall effect. However, this local configuration is prone to spurious signals. In this work, we address spurious Hall effects that can contaminate the spin Hall signal in these FM/HM T-shaped nanostructures. The most pronounced Hall effects in our $Co_{50}Fe_{50}$/Pt nanostructures are the planar Hall effect and the anomalous Hall effect generated in the FM nanowire. We find that the planar Hall effect, induced by misalignment between magnetization and current direction in the FM wire, is manifested as a shift in the measured baseline resistance, but does not alter the spin Hall signal. In contrast, the anomalous Hall effect, arising from the charge current distribution within the FM, adds to the spin Hall signal. However, the effect can be made insignificant by minimizing the shunting effect via proper design of the device. We conclude that local spin injection in FM/HM nanostructures is a suitable tool for measuring spin Hall signals and, therefore, a valid method for magnetic state readout in prospective spin-based logic.

## I. INTRODUCTION

Spin-orbitronics[1] is a new field in condensed matter physics that aims to utilize different phenomena in magnetism and spintronics caused by the spin-orbit coupling (SOC).[2–5] One of the most studied phenomena in spin-orbitronics is the spin Hall effect (SHE), occurring in heavy metals (HM) with a strong bulk SOC such as Pt, Ta and W.[6–11] The SHE allows for interconversion between charge currents and spin currents[6] and has a strong potential to be harnessed for energy-efficient logic and memory tasks for processing of information.[5,12–15] The SHE converts a charge current into a spin current and can be used to write magnetic memory states.[16,17] The reciprocal effect (inverse SHE, ISHE), responsible for the conversion of a spin current into a charge current, can achieve reading of magnetic memory states.[13,18] Particularly, the readout of non-volatile magnetic states via the inverse SHE is one of the main components of the magneto-electric spin-orbit (MESO) logic, because it generates an electromotive force (voltage) and a current that can be used to cascade different elements of the logic-built-in-memory architecture. In contrast, the magnetic state readout by magnetoresistance techniques[19,20], widely used in our modern technology, do not have the ability to drive other circuit elements. The MESO logic is computed to have the lowest energy consumption per 32-bit arithmetic logic unit in comparison to other beyond-CMOS logic proposals.[13]

A well-established device to study the (I)SHE is the spin absorption in a lateral spin valve (LSV) where, with a nonlocal configuration, a pure spin current is injected from a ferromagnetic (FM) electrode into a nonmagnetic transport channel and detected in a spin-orbit material (SOM) electrode.[7,8,11,21,22] Using pure spin currents is convenient for the quantification of the SHE, as it eliminates spurious signals associated to local effects. In contrast, for a potential application such as MESO logic, LSVs present the disadvantage of a strong reduction of the spin signal caused by the exponential decay of the spin current and the shunting of the generated charge current in the transport channel,[23] as well as the spin backflow in the FM electrode. A way to overcome these issues is by removing the transport channel and inject and detect the spin current directly at the FM/SOM interface. Such simpler FM/SOM T-shaped nanostructures have been used to study the spin properties of HMs[10,18,24–26] and topological insulators[27] and can induce spin Hall signals three orders of magnitude larger in comparison to signals in LSVs. The spin Hall signals are easily measurable at room temperature and, furthermore, such device configuration allows for independent scalable readout in terms of voltage and current,[18] which make it ideal for the magnetic state readout in the MESO logic device. However, the local electronic measurement configuration in these devices, exploiting spin-polarized currents instead of pure spin currents, might give rise to spurious transverse voltages. The Hall effects that strongly emulate the (I)SHE are the anomalous Hall effect (AHE)[28] and the planar Hall effect (PHE).[29] The understanding of the different spurious effects in the FM/HM T-shaped nanostructures, which allows for disentanglement of the proper SHE, is relevant for the reliability of the magnetic state readout and the realization of the MESO logic device.

In this paper, we disentangle the Hall effects that could contaminate the spin Hall signal when measuring the ISHE in FM/HM nanostructures due to (1) misalignment of the magnetization and the charge current in the planar direction and (2) vertical lines in the inhomogeneous charge current density distribution at the FM side of the injection region. The former induces the PHE in the FM, whereas the latter can lead to the AHE in the FM. We identify that the PHE induces a shift of the baseline of the transverse resistance and distort its shape, but does not affect the magnitude of the spin Hall signal, that is, the difference in the resistance between the two magnetization directions of the FM. In contrast to the PHE case, the AHE appears with the same symmetry as the ISHE, and therefore disentanglement from the spin Hall signal is not straightforward. However, by combining electrical measurements and a three-dimensional finite-element-method (3D FEM) simulation, the AHE contribution to the measured signal can be estimated. Further modelling shows that the AHE contribution can be minimized by tuning the thicknesses of the FM and HM electrode in the nanostructure. We find that the AHE accounts for less than 10% of the measured signal for the $Co_{50}Fe_{50}$ (15 nm)/Pt (8 nm) sample used here. Our results show that spurious Hall effects in FM/HM T-shaped nanostructures can be distinguished and minimized. Therefore, these devices can be used as a simple tool to measure the spin Hall signal in order to extract the spin-to-charge conversion efficiency of the system, as well as a reliable method for the readout of in-plane magnetic states.[18]

## II. EXPERIMENTAL DETAILS

The FM/HM nanostructures consist of a T-shaped HM nanostructure and an FM electrode where the tip of the FM nanowire is overlapping with the intersection of the T-shaped nanostructure, see Figs. 1(a) and 1(b) for a scanning electron microscopy (SEM) image of the device. In this study, we will use Pt as the HM and $Co_{50}Fe_{50}$ (from here on named CoFe) as the FM material. The devices are fabricated on $SiO_2$(150 nm)/Si substrates in two steps, each step involving electron-beam lithography (eBL), metal deposition, and lift-off process. The first step defines the T-shaped nanostructure by eBL, followed by Pt deposition via magnetron sputtering (1.3 Å/s, $p_{Ar} = 3$ mTorr, $P = 80$ W, $p_{Base} \sim 5 \times 10^{-8}$ mTorr at room temperature). After lift-off, Ar-ion beam milling is performed at grazing incidence to remove side walls on the Pt electrode. In a second eBL step, the FM nanowire is patterned. Before CoFe deposition, an Ar-ion beam milling is performed at normal incidence to clean the Pt surface and guarantee a highly transparent interface between the Pt and CoFe electrodes. Lastly, the CoFe is deposited by magnetron sputtering (0.24 Å/s, $p_{Ar} = 2$ mTorr, $P = 30$ W, $p_{Base} \sim 5 \times 10^{-8}$ mTorr at room temperature) and subsequently lifted-off in acetone.

The width and thickness of the Pt and CoFe electrodes for the device presented in this work are $w_{CoFe} = 185$ nm, $t_{CoFe} = 15$ nm, $w_{Pt} = 215$ nm, and $t_{Pt} = 8$ nm. The resistivities are measured to be $\rho_{CoFe} = 91$ μΩcm and $\rho_{Pt} = 154$ μΩcm. The electrical transport measurements are performed in a Physical Property Measurement System from Quantum Design, where we apply an in-plane magnetic field with a superconducting solenoid magnet at 300 K. The measurements are executed using the 'DC reversal technique' with a Keithley 2182 nanovoltmeter and 6221 current source.

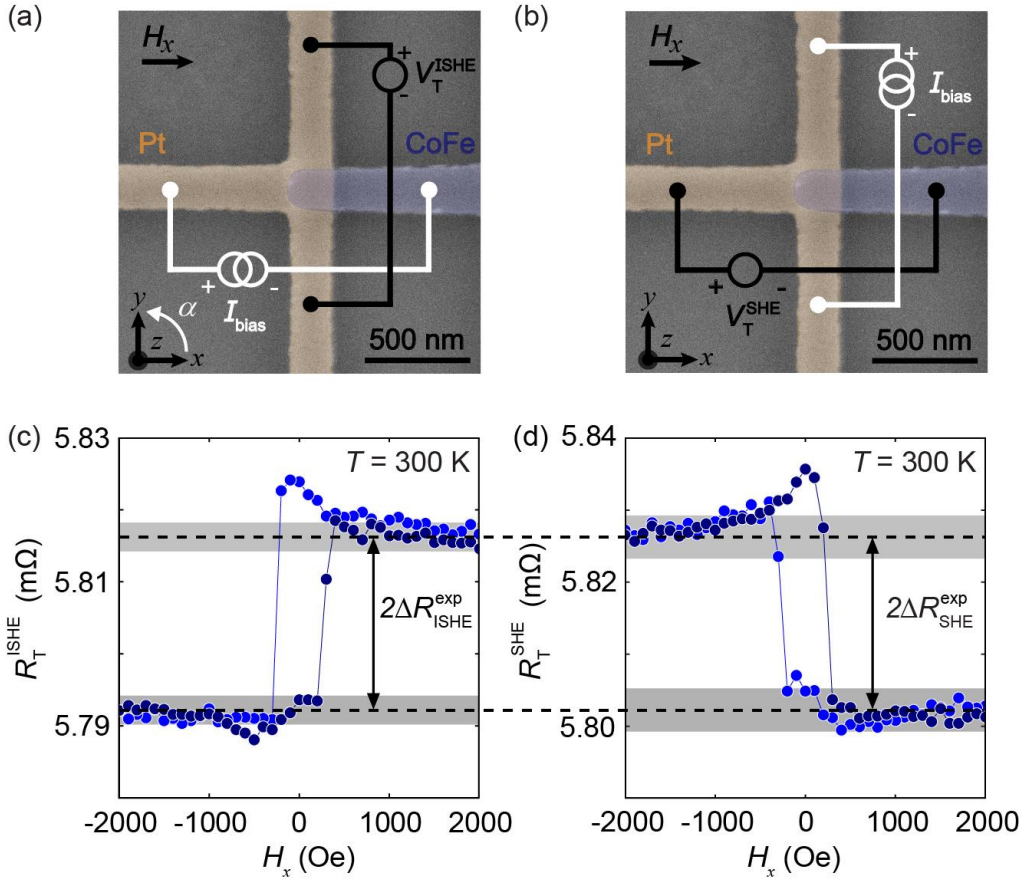

FIG. 1: (a) and (b) False-colored top view SEM image of an FM/HM nanostructure with the ISHE and SHE measurement configuration, respectively, and the orientation of the external magnetic field $H_x$. Blue indicates the CoFe electrode and yellow marks the T-shaped Pt nanostructure. (c) and (d) Evolution of the transverse resistance ($R_T^{ISHE}$ in the ISHE configuration and $R_T^{SHE}$ in the SHE configuration, respectively) as a function of $H_x$ (trace and retrace) measured at $I_{bias}$= 50 µA and 300 K. The difference between the low- and high-resistance states (dashed black lines; the grey shaded area represents the associated error) is the experimentally obtained spin Hall signal $2\Delta R_{(I)SHE}^{exp}$.

## III. RESULTS AND DISCUSSION

### A. Spin Hall effect and inverse spin Hall effect

Figures 1(a) and 1(b) show a top-view SEM image of the device with the configuration of the ISHE and SHE measurement setups, respectively, where the magnetization $\boldsymbol{m}$ of the CoFe electrode is aligned along the easy axis and can be switched with an external magnetic field $H_x$. The ISHE is measured by applying a bias current $I_{bias}$ from the CoFe electrode into the Pt electrode, which injects a spin current into Pt, polarized in the $x$-direction. The strong SOC in Pt allows for the ISHE to produce a transverse charge current, which can be detected as a voltage $V_T^{ISHE}$ in open circuit conditions along the transverse Pt electrode. The reversal of $\boldsymbol{m}$ induces a sign change of $V_T^{ISHE}$. In the SHE configuration, $I_{bias}$ is applied through the transverse Pt electrode such that a spin current polarized in the $x$-direction is generated via SHE, resulting in spin accumulation at the surfaces of the Pt. The top surface can be probed with the FM electrode as the Fermi level of the CoFe electrode aligns with the electrochemical potential of the majority (minority) spins of the spin accumulation and a positive (negative) interface voltage $V_T^{SHE}$ is created when $\boldsymbol{m}$ is oriented along $+x$ ($-x$).[10,18] The spin Hall resistance is defined as $R_T^{(I)SHE} = V_T^{(I)SHE}/I_{bias}$. The two resistance states can be associated with the magnetic state of the ferromagnet (i.e., we are reading out the magnetic state) and the difference between these two resistance states is the spin Hall signal $2\Delta R_{(I)SHE}$.

Figures 1(c) and 1(d) present the transverse resistance $R_T$ as a function of $H_x$ for the ISHE and SHE measurement configurations, respectively. Figure 1c shows that, as long as the magnetization is aligned along the $+x$-direction, the ISHE induces a constant $R_T^{ISHE}$, the high resistance state. The reversal of the magnetization to $-x$-direction by sweeping $H_x$ switches $R_T^{ISHE}$ to a constant low resistance state. The same behavior is observed when sweeping $H_x$ in opposite direction, with a hysteresis associated to the magnetization of the CoFe electrode aligned along its easy axis. If we assume that the overall observed signal is purely coming from the ISHE, we extract a spin Hall signal $2\Delta R_{ISHE}^{exp} = (24 \pm 2)$ mΩ. Figure 1(d) presents the transverse resistance $R_T^{SHE}$ for the SHE measurement configuration, which gives $2\Delta R_{SHE}^{exp} = (24 \pm 3)$ mΩ. The spin Hall signals of the ISHE and the SHE are the same, as expected from the Onsager reciprocity.[21] The study in the remainder of this paper is focused on the ISHE, but the results are also valid for the SHE (see Appendix 1).

The ISHE in FM/HM nanostructures can be described by the one-dimensional (1D) spin diffusion model. Note that, in this work, we study the CoFe/Pt system, but the model can be extended to other FM/HM systems with transparent interfaces. If the thicknesses of the Pt and CoFe electrodes are much larger than the corresponding spin diffusion lengths, the spin Hall signal is given by $\Delta R_{(I)SHE} = G \times \lambda_{eff}$, with "geometrical factor" $G = 1/((t_{CoFe}/\rho_{CoFe} + t_{Pt}/\rho_{Pt})w_{Pt})$ and "efficiency factor" $\lambda_{eff} = P_{CoFe}\theta_{SH}\lambda_{Pt}/(1 + \lambda_{Pt}\rho_{Pt}/\lambda_{CoFe}\rho^*_{CoFe})$ (Ref. [18]). Here, $P_{CoFe}$ is the polarization of CoFe and $\rho^*_{CoFe} = \rho_{CoFe}(1 - P^2_{CoFe})$. $\theta_{SH}$ is the spin Hall angle of Pt, and $\lambda_{Pt}$ and $\lambda_{CoFe}$ are the spin diffusion lengths of Pt and CoFe, respectively. The efficiency of the materials (FM/HM) system, given by $\lambda_{eff}$, is independent on the geometry and the resistivities of the materials. We find that, for the CoFe/Pt nanostructure under study, $G = (220 \pm 20)$ mΩ/nm. By combining $G$ with the experimentally obtained spin Hall signal $2\Delta R^{exp}_{ISHE} = (24 \pm 2)$ mΩ, the efficiency of this CoFe/Pt nanostructure is estimated to be $\lambda_{eff} = (0.055 \pm 0.007)$ nm, in good agreement with the efficiency for this materials system determined by an extensive study on the scalability of the local devices which showed that $\lambda^{CoFe/Pt}_{eff} = (0.05 \pm 0.01)$ nm (Ref. [18]).

Even though the efficiency of the CoFe/Pt system in this nanostructure is the same as the expected efficiency presented in Ref. 13, it is of paramount importance to rule out possible spurious effects in the spin Hall signal. Since the spin Hall resistance is a transverse signal showing the hysteresis of the FM element, any contamination in the measurement must arise from other Hall effects present in the FM. The Hall effects with the most significant magnitudes are the PHE and the AHE. These originate in CoFe wire, and may appear in the transverse signal due to electrical shunting through Pt. The contribution of these Hall effects will depend mainly on the resistivities and thicknesses of the Pt and CoFe electrodes and the interface resistance. Although the ordinary Hall effect (OHE) in Pt, in conjunction with out-of-plane fringe fields from the FM electrode tip, is also a potential source for a spurious signal, it is too weak to account for any measured signal (see Appendix 2).

### B. Planar Hall effect

It is known that ferromagnetic *3d* metals and alloys possess a strong anisotropic magnetoresistance effect (AMR) due to *s–d* scattering processes from the conduction *s* state to the localized *d* states. AMR is manifested as the difference in the longitudinal resistance as a response to a parallel or perpendicular orientation of $\boldsymbol{m}$ with respect to the applied charge-current density $\boldsymbol{j}_c$ giving rise to two distinct resistivities $\rho_\parallel$ and $\rho_\perp$, respectively.[30,31] The same phenomenon induces a transverse resistance, known as the PHE. The behavior of the planar Hall resistivity as a function of the angle $\varphi$ between an in-plane magnetization $\boldsymbol{m} = (m\cos(\varphi), m\sin(\varphi), 0)$ and $\boldsymbol{j}_c$ is [32,33]

$$\rho^{PHE}_{xy}(\varphi) = (\rho_\parallel - \rho_\perp)\cos(\varphi)\sin(\varphi) = \frac{(\rho_\parallel - \rho_\perp)}{2}\sin(2\varphi). \tag{1}$$

This means that PHE does not influence the spin Hall signal in our device except in the presence of a finite in-plane angle between $I_{bias}$ and $\boldsymbol{m}$ of the CoFe electrode, as show in Fig. 2(a). This could occur due to an angular misalignment in the experimental setup and/or in the nanofabrication.

To disentangle the PHE from the obtained spin Hall signal, the measurement configuration of the ISHE measurement setup [Fig. 1(a)] is used with the external magnetic field fixed to 5 kOe while an in-plane rotation $\alpha$ of the device is performed. Since this field is large enough to overcome the shape anisotropy of the CoFe electrode, **m** will align with the field direction. In the measurement setup, an ISHE contribution will arise from the spin current polarized along the *x*-direction, which is proportional to $m \cos(\varphi)$. The contribution of the PHE will be proportional to $\sin(2\varphi)$ [Eq. (1)]. Hence, the dependence of total transverse resistance dependence on the rotation angle $\alpha$ is described by:

$$R_T^{ISHE} = R_{ISHE} + R_{PHE} + b = a_{ISHE} \cos(\alpha + \alpha_0) + a_{PHE} \sin(2(\alpha + \alpha_0)) + b, \quad (2)$$

where $a_{ISHE}$ and $a_{PHE}$ are the amplitudes of the ISHE and PHE, respectively, $\alpha_0$ is a misalignment angle, and $b$ is the baseline resistance of the measurement. The ISHE and the PHE can thus be identified from the difference in the angular dependent behavior of the transverse resistances, $R_{ISHE}$ and $R_{PHE}$, respectively.

Figure 2(b) shows the experimental transverse resistance as a function of $\alpha$ (black solid dots) together with a fit of the data to Eq. (2) (dashed red curve). The obtained fitting parameters are $a_{ISHE} = (11.8 \pm 0.7)$ m$\Omega$, $a_{PHE} = -(99.3 \pm 0.7)$ m$\Omega$, $\alpha_0 = -(1.0 \pm 0.2)°$ and $b = (5.8001 \pm 0.0005)$ $\Omega$. The angle $\alpha_0$ has a small and realistic value for the misalignment between the applied magnetic field and the CoFe nanowire that depends on the placement of the sample in the measurement setup. The contributions of the ISHE and PHE to the fit are as well presented in Fig. 2(b) by the blue and cyan curve, respectively, showing that the PHE resistance has a much higher amplitude than the ISHE resistance. Figure 2(c) displays a zoom of the ISHE contribution from where we identify the PHE resistance $R_{PHE}^{0°} = R_{PHE}^{180°} = (3.5 \pm 0.7)$ m$\Omega$ due to the presence of the misalignment, whereas the ISHE resistance is $R_{ISHE}^{0°} = (11.8 \pm 0.7)$ m$\Omega$ and $R_{ISHE}^{180°} = -(11.8 \pm 0.7)$ m$\Omega$. The cyan dashed line represents the contribution of the PHE when $\alpha$ is 0° and 180° ($R_{PHE}^{0°/180°}$). The black dashed lines depict the difference between $R_{ISHE}^{0°}$ and $R_{ISHE}^{180°}$, *i.e.*, the expected spin Hall signal from the fit $2\Delta R_{ISHE}^{fit}$, which should correspond to $2\Delta R_{ISHE}^{exp}$ measured in the magnetic-field dependence. Appendix 1 presents the analysis for the SHE configuration, showing the same behavior for the PHE resistance.

Figure 2(d) shows the magnetic field dependence of $R_T^{ISHE}$, whose raw date is plotted in Fig. 1(c), after the baseline $b$ obtained from the fit is subtracted. The positive and negative externally applied magnetic field correspond to the angles 0° and 180° in the angle dependence [Fig. 2(c)], respectively. The value of $2\Delta R_{ISHE}^{exp} = (24 \pm 2)$ m$\Omega$, as determined by the magnetic field dependence at high magnetic field, is observed to be the same as obtained from the angle dependence [$2\Delta R_{ISHE}^{fit} = (24 \pm 1)$ m$\Omega$], but shifted by the PHE an amount of $R_{PHE}^{0°/180°}$. Hence, at the negative and positive saturation fields, the low and high resistive states are given by ($-R_{ISHE} + R_{PHE}$) and ($R_{ISHE} + R_{PHE}$) for a negative misalignment angle. In the case $\alpha_0$ were positive in this configuration, the PHE resistance would be negative and would shift the transverse resistance down (for more details, see Appendix 3). We note that, as the PHE has the same contribution ($+R_{PHE}$) for the negative and positive magnetic fields, $R_T^{ISHE}$ is shifted but the extraction of $2\Delta R_{ISHE}$ is not contaminated by the PHE.

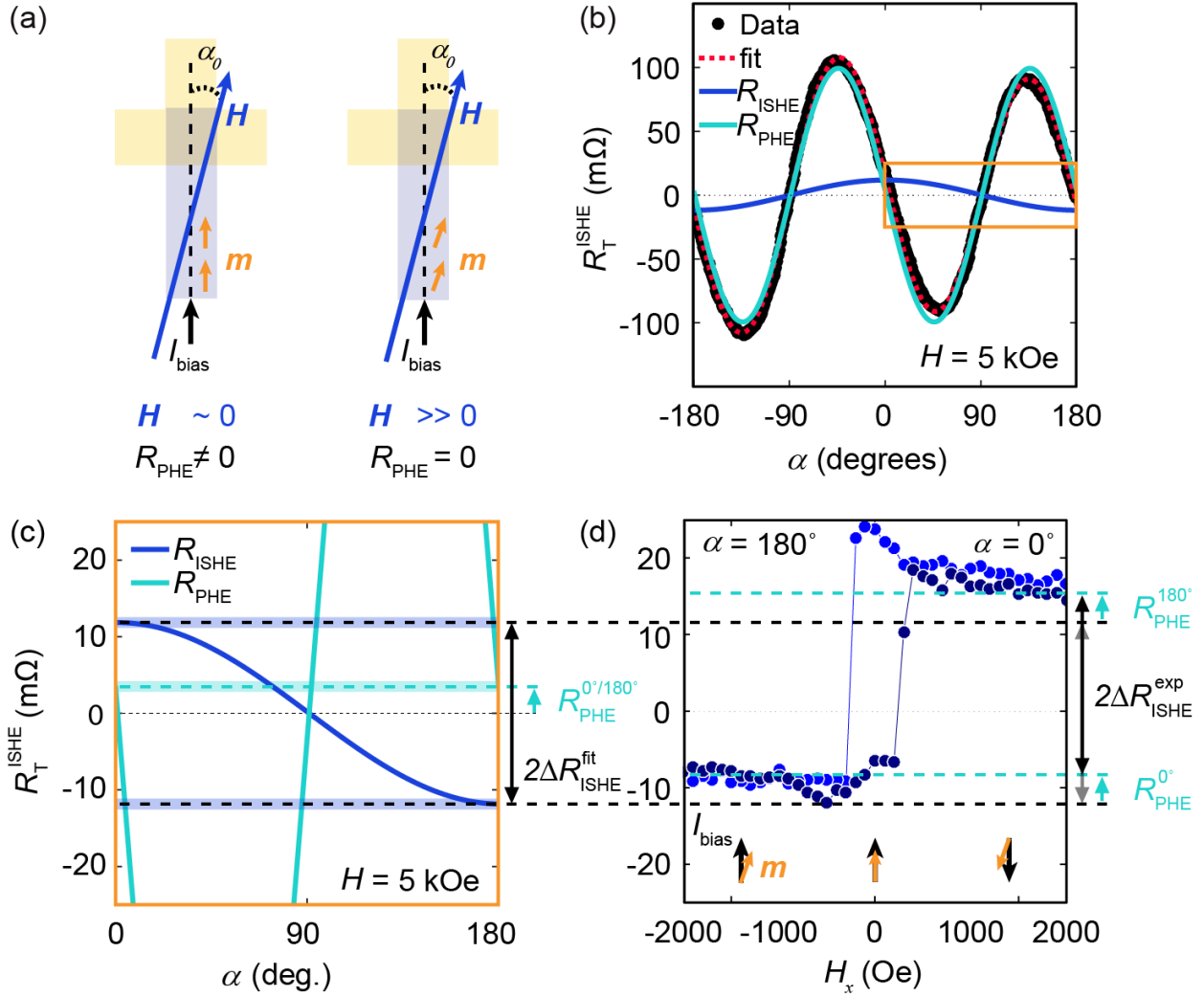

FIG 2: (a) A representation of an FM electrode with a magnetization $m$ and the effect on $m$ of a misalignment angle $\alpha_0$ between the applied current bias $I_{bias}$ and an external in-plane magnetic field $H$. Left panel: At small $H$, $m$ aligns with the easy axis (i.e. the longest dimension of the FM electrode), which is parallel to $I_{bias}$, such that $\varphi = 0$ and the PHE contribution is zero. Right panel: At large $H$, $m$ aligns with $H$, such that $\varphi = \alpha_0$ and the PHE contribution is nonzero. (b) The transverse resistance $R_T^{ISHE}(\alpha)$ as a function of the in-plane angle $\alpha$ (Fig. 1a) at $I_{bias}$= 50 μA and 300 K with a fixed magnetic field of 5 kOe. The black solid dots are the measured data whereas the red dashed line is a fit to Eq. 2. The fit can be separated into an ISHE component and a PHE component shown by the blue and cyan curve, respectively. (c) A zoom of the ISHE component presented in panel b indicated by the orange square. The cyan dashed line presents the magnitude of the PHE signal at $\alpha = 0°$ and $\alpha = 180°$ ($R_{PHE}^{0°/180°}$). (d) The transverse resistance $R_T^{ISHE}$ versus the applied magnetic field in the x-direction as given in Fig. 1c after subtraction of the baseline obtained from the fit ($b = 5.8001\ \Omega$). The high resistance state corresponds to $\alpha = 0°$ whereas the low resistance state corresponds to $\alpha = 180°$. A shift of the transverse resistance by an amount of $R_{PHE}^{0°/180°}$ (cyan-dashed line) is observed at large magnetic fields where $m$ and $I_{bias}$ are misaligned. The experimental spin Hall signal $2\Delta R_{ISHE}^{exp}$ is the same as the spin Hall signal obtained from the fit on the angle dependence of the transverse resistance $2\Delta R_{ISHE}^{fit}$.

## C. Anomalous Hall effect

The AHE is the transverse deflection of charge carriers leading to a transverse voltage observed in materials with a net magnetization when a charge current is applied.[28] This effect is generally measured in a Hall cross with an out-of-plane magnetic field ($H_z$). Although, in our measurement configuration, the applied magnetic field in the FM/HM nanostructures is in plane ($H_x$), the AHE can be present due to an inhomogeneous distribution of the charge current density near the CoFe/Pt interface.

Unlike the case of the PHE, the symmetry of the AHE and the SHE is the same, and thus the two contributions cannot be disentangled with an angular-dependent measurement. Therefore, 3D FEM simulations, within the framework of the two-current drift-diffusion model,[34,35] are performed to retrieve the ISHE and AHE contributions to the total transverse resistance measured in the ISHE configuration ($2\Delta R_{\text{ISHE}}^{\text{exp}}$, see Figs. 1a and 1c). The geometrical construction and 3D-mesh [Fig. 3(a)] were elaborated using the free software GMSH[36] with the associated solver GETDP[37] for calculations, post-processing and data flow controlling. The top CoFe/Pt interface is assumed to be transparent with no spin memory loss and the lateral CoFe/Pt interface is considered to be insulating as the lateral side of the Pt is not etched.[10,18] Furthermore, the spin polarization of CoFe and the spin Hall angle of Pt are set to $P_{\text{CoFe}} = 0.48$ (Ref. [38]) and $\theta_{\text{SH}} = 0.27$ (Ref. 7), respectively. We assume that $\rho_{\text{Pt}}\lambda_{\text{Pt}} = 0.77$ fΩ m² (Ref. [39]) and $\rho_{\text{CoFe}}\lambda_{\text{CoFe}} = 1.29$ fΩ m² (Ref. [38]) such that, considering the resistivities of our Pt and CoFe electrodes, the spin diffusion lengths are $\lambda_{\text{Pt}} = 0.5$ nm and $\lambda_{\text{CoFe}} = 1.4$ nm.

Figure 3(a) shows the electric potential in the FM/HM nanostructure when applying $I_{\text{bias}}$ from port 1 to port 2 of the model. The inset of Fig. 3(a) shows a side view of our nanostructure with the distribution and orientation of the charge current density in the CoFe/Pt interface region. The charge current density possesses a component in the *z*-direction inside the CoFe electrode, perpendicular to the in-plane magnetization (*x*-direction). This can induce a transverse voltage in the *y*-direction due to AHE with the same symmetry in the magnetic hysteresis loop as the ISHE. The transverse signal is the difference in electric potential between port 3 and port 4, which is normalized by the applied current.

The anomalous Hall angle $\theta_{\text{AH}} = \rho_{\text{AH}}/\rho_{\text{CoFe}}$, which quantifies the strength of the AHE in materials, serves as an input parameter for the 3D FEM and is experimentally obtained using a standard Hall cross (HC) measurement. The inset of Fig. 3(b) presents an SEM image of a CoFe Hall cross, next to our CoFe/Pt nanostructure (Figs. 1(a) and 1(b)), with the measurement configuration. A transverse voltage ($V_{xy}^{\text{HC}}$) appears when applying a bias current $I_{\text{bias}}$ along the horizontal electrode and an external out-of-plane magnetic field ($H_z$). Figure 3(b) plots the transverse resistivity, $\rho_{xy}^{\text{HC}} = (V_{xy}^{\text{HC}}/I_{\text{bias}})t_{\text{CoFe}}$, as a function of $H_z$. The dashed lines are linear fits to the transverse resistivity at large magnetic field, corresponding to the ordinary Hall contribution, and the difference between the zero field extrapolations of the two fits is twice the anomalous Hall resistivity, $2\rho_{\text{AHE}}$. We find that the anomalous Hall resistivity of the CoFe in this nanostructure is $\rho_{\text{AHE}} = (0.562 \pm 0.001)$ µΩcm such that the anomalous Hall angle yields $(0.618 \pm 0.001)\%$.

This experimental $\theta_{\text{AH}}$ is implemented in the 3D FEM to extract the AHE contribution to the transverse signal. We define $2\Delta R_{\text{AHE}}$ as the difference between the transverse resistance induced by the AHE in the ISHE configuration at positive and negative saturated magnetization. As the magnetic hysteresis loops of the ISHE and the AHE are equal, the transverse resistance is the addition of the two contributions, i.e., $2\Delta R_{\text{ISHE}}^{\text{exp}} = 2\Delta R_{\text{ISHE}} + 2\Delta R_{\text{AHE}}$. Finally, from the 3D FEM, we extract the spin Hall signal and the AHE contribution to be $2\Delta R_{\text{ISHE}} = (18\pm2)$ mΩ and $2\Delta R_{\text{AHE}} = 2.1$ mΩ, respectively. According to the 3D FEM, we should thus measure $2\Delta R_{\text{ISHE}}^{\text{exp}} = (20 \pm 2)$ mΩ. The FM/HM interface in our nanostructures might not be as perfectly transparent as assumed in the 3D FEM simulation, this may lead to a small increase of the experimental spin Hall signal, but indeed, the spin Hall signal obtained from the 3D FEM is, within the error, in agreement with the spin Hall signal that was experimentally measured [$2\Delta R_{\text{ISHE}}^{\text{exp}} = (24\pm2)$ mΩ].

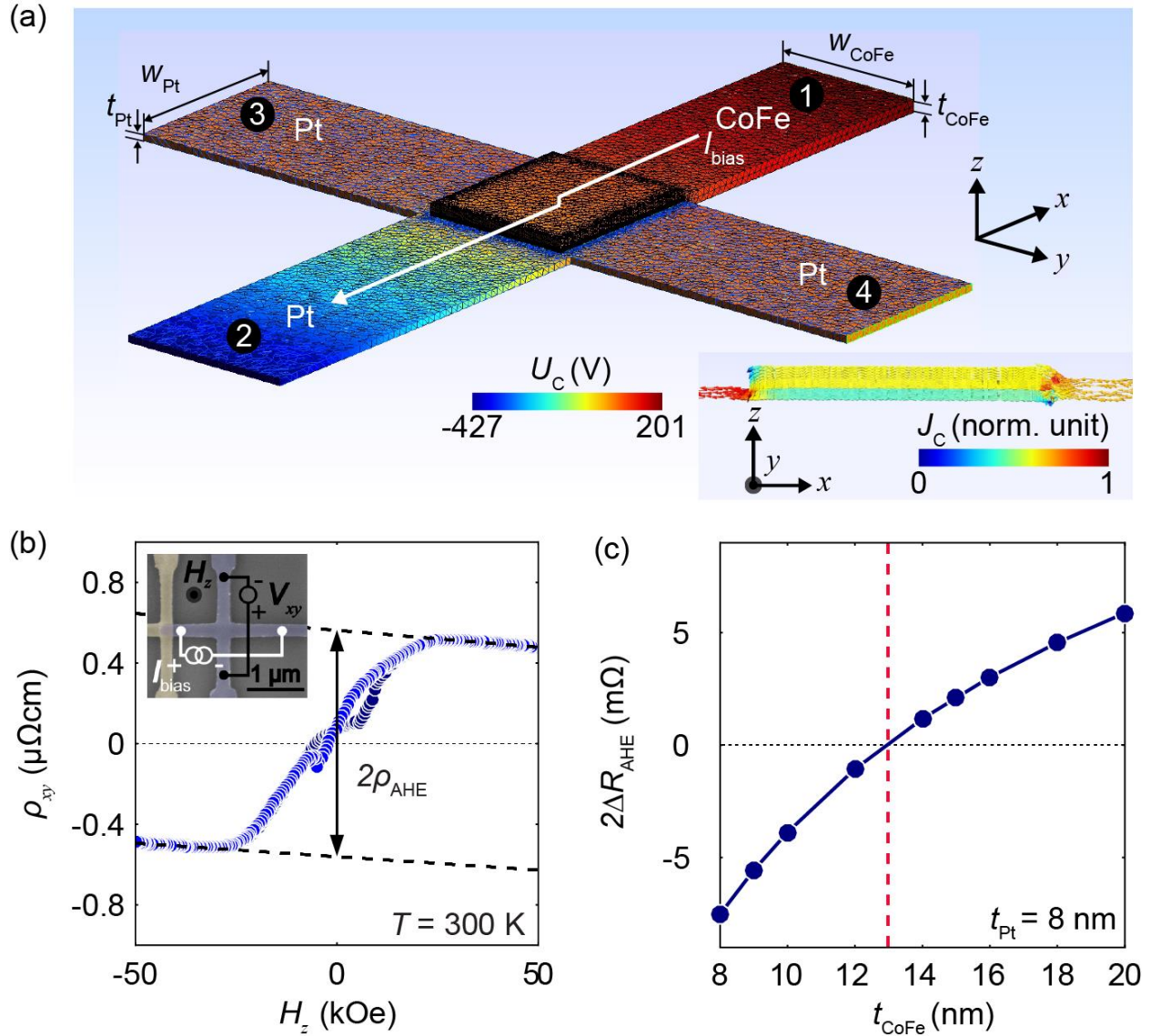

FIG 3: (a) Geometry, mesh and the electric potential map of the 3D FEM model used for the FM/HM nanostructure to estimate the AHE in the ISHE configuration. The 3D color map of electrical potential induced by a bias current $I_{bias}$ (white arrow) forced from the CoFe electrode (port 1) to the Pt electrode (port 2). As all the equations used for the simulations are linear, the voltage values correspond to an injected current arbitrarily set at 1 A. Inset: side view of the device (xz-plane) picturing the charge current density

lines. It shows the existence of a *z*-component in the charge current density which gives rise to the AHE in the CoFe electrode. (b) The out-of-plane magnetic field dependence of the transverse resistivity in a CoFe cross-shaped nanostructure. The anomalous Hall resistivity $\rho_{AHE}$ extracted from this measurement is used to determine the strength of the AHE. Inset: AHE measurement setup with the CoFe Hall cross depicted in blue. (c) The CoFe nanowire-thickness dependence of the AHE signal obtained using 3D FEM simulations. The line is a guide to the eye. The AHE can be eliminated by optimizing the thickness of the CoFe nanowire. The optimum CoFe thickness at which the AHE contribution vanished is ∼13 nm (red dashed line) in this device with a Pt nanowire thickness of 8 nm and the resistivities as presented in the text.

Finally, from the simulations we obtain that the AHE contribution to the total spin Hall signal is about 10% for the CoFe/Pt nanostructure under study. However, this contamination can be reduced by adjusting the thickness of the CoFe nanowire. The magnitude of the *z*-component of charge current density in CoFe depends on the thicknesses of the CoFe electrode and the Pt nanowire. 3D FEM simulations, as discussed above, are performed for various CoFe thicknesses to extract the dependence of the AHE signal considering the other parameters in the model system to be unchanged. Figure 3c shows that the AHE contribution to the measured signal can be positive or negative, depending on the thickness of the CoFe nanowire. This can be understood considering the modeled charge-current-density distribution $j_c$ presented in the inset of Fig. 3(a). At the tip of the CoFe electrode (left side), $j_c$ is observed to have components oriented in −*z*-direction while in the area where the CoFe electrode starts to overlap the Pt electrode (right side), $j_c$ is oriented in the +*z*-direction. The two orientations of the charge current density induce opposite AHE resistances that compete with one another. The charge-current-density distribution at the interface between the Pt and CoFe electrode depends on the thicknesses of both electrodes, and thus the two AHE contribution can be compensated by a suitable device geometry. The contribution of the AHE, according to our simulations, is eliminated when the CoFe thickness is about 13 nm for an 8-nm-thick Pt nanowire (red dashed line). Choosing the thickness of the FM electrode to be about one-and-a-half times the thickness of the HM electrode ($t_{FM} \sim 1.5\ t_{HM}$) can be taken as a rule of thumb to minimize the AHE when the FM and HM resistivities are of the same order.

The different Hall effects discussed here for CoFe and Pt are present in any metallic FM (PHE and AHE) and HM (OHE), therefore the method to disentangle these Hall effects shown in this work can be applied to any other metallic FM/HM system with a transparent interface.

## IV. CONCLUSIONS

In conclusion, we have studied the appearance of spurious Hall effects in the local CoFe/Pt T-shaped nanostructures that are promising for magnetic state readout via (I)SHE measurements. The strongest Hall effects are induced in the FM electrode (PHE and AHE) and transferred into the HM electrode, whereas the OHE in the HM caused by the stray fields of the FM electrode is negligible. The PHE appears with a different symmetry than the ISHE such that an angle misalignment between the magnetic field and the FM electrode can induce a shift in the transverse resistance. This PHE shift can be obtained and corrected for by performing a full angular-dependent measurement of the transverse resistance. The PHE contribution, however, does not affect the reading of the spin Hall signal at saturated magnetic fields. The AHE, however, appears with the same symmetry as the (I)SHE in the measurement. These two contributions can be disentangled by 3D FEM simulations. We observe an AHE contamination of 10% of the transverse resistance in our sample. Further investigation shows that optimizing the thickness of the CoFe electrode with respect to the Pt electrode minimizes of the AHE contribution. As the one-dimensional spin diffusion model accounts for any FM/HM system and the PHE and AHE are valid for all metallic FM, the results can be generalized to any metallic FM/HM T-shaped device. Finally, we emphasize that the FM/HM T-shaped nanostructures are an appropriate tool to measure the (I)SHE in which parasitic effects can be eliminated by proper alignment and optimized design of the nanostructures and are thus a viable option for future energy-efficient spin-based logic technology.

## Acknowledgments

This work is supported by Intel Corporation through the Semiconductor Research Corporation under MSR-INTEL TASK 2017-IN-2744 and the 'FEINMAN' Intel Science Technology Center, and by the Spanish MICINN under the Maria de Maeztu Units of Excellence Programme (MDM-2016-0618) and under project numbers MAT2015-65159-R and RTI2018-094861-B-100. V.T.P. acknowledges postdoctoral fellowship support from 'Juan de la Cierva—Formación' programme by the Spanish MICINN (grant number FJCI-2017-34494). N.L. acknowledges funding from the European Union's Horizon 2020 research and innovation program under the Marie Słodowska Curie Grant Agreement No.~844304 (LICONAMCO).

## Appendix 1: The PHE contamination in the SHE measurement configuration

The main text discusses how the PHE can be separated from the ISHE measurement in our CoFe/Pt nanostructures by analyzing the in-plane angle dependence of the transverse resistance. The same analysis holds true for the SHE measurement because of the Onsager reciprocity between the ISHE and the SHE [Ref. [21]]. Figure 4(a) presents a false-colored SEM image showing the Pt and CoFe electrodes as the yellow and blue areas, respectively, and the SHE measurement configuration. Figure 4(b) graphs the in-plane angle ($\alpha$) dependence of the transverse resistance measured in this configuration ($R_\text{T}^\text{SHE}$). The black solid dots are the experimental data, and the dashed red line is the fit to Eq. (2). The fitting parameters are $a_\text{ISHE} = -(10.3 \pm 0.8)$ mΩ, $a_\text{PHE} = -(99.4 \pm 0.8)$ mΩ, $\alpha_0 = -(0.9 \pm 0.2)$ °, and $b = -(5.8121 \pm 0.0005)$ Ω. These are comparable to the fitting parameters obtained from the ISHE measurement in the main text. The fit can be decomposed into a SHE and a PHE component due to the difference in the angle dependence being $\cos(\varphi)$ [blue line] and $\sin(2\varphi)$ [cyan line], respectively.

Figure 4(c) displays a zoom of the area of interest, being the orange box in Fig. 4b, to compare the angular dependence with the magnetic field dependence [Fig. 4(d)] of the transverse resistance. We extract the spin Hall signal to be $2\Delta R_\text{SHE}^\text{fit} = (21\pm1)$ mΩ and $2\Delta R_\text{SHE}^\text{exp} = (24\pm3)$ mΩ, meaning that they are the same within the error bar. The PHE signal is $R_\text{PHE}^{0°/180°} = (3.1\pm0.8)$ mΩ for the SHE measurement set-up. Finally, we can conclude that the fit correctly predicts the spin Hall signal and that the PHE induces a shift in the base line, equivalent to the results shows in the main text for the ISHE.

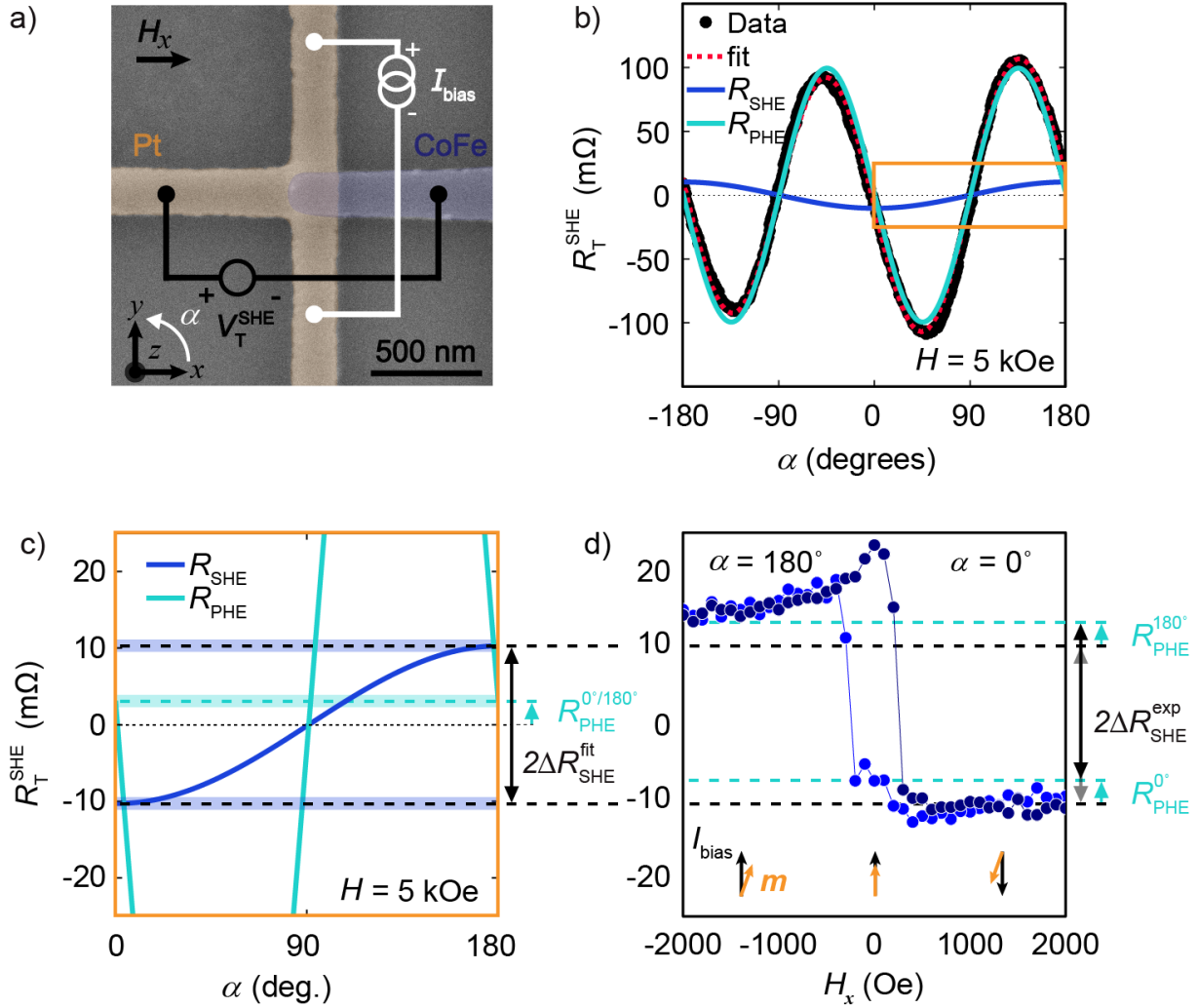

FIG. 4: (a) False-colored top view SEM image of the FM/HM nanostructure where the FM (CoFe) and HM (Pt) electrodes are indicated by blue and yellow, respectively. The SHE measurement configuration is displayed. The in-plane rotation of the magnetic field is described by the angle α. (b) The transverse resistance $R_T^{SHE}$ as a function of $\alpha$ at 300 K with a fixed magnetic field of 5 kOe. The black dots are the measured data, and the red dashed line is a fit to Eq. 2. The fit is separated into an SHE component and a PHE component shown by the blue and cyan curve, respectively. (c) A zoom of ISHE component presented in panel b indicated by the orange square. The cyan-dashed line presents the magnitude of the PHE signal at $\alpha = 0°$ and $\alpha = 180°$ ($R_{PHE}^{0°/180°}$). (d) The transverse resistance $R_T^{SHE}$ versus the applied magnetic field in the x-direction as given in Fig. 1d after subtraction of the baseline obtained from the fit ($b = 5.8121\ \Omega$). The low resistance state corresponds to α = 0° whereas the high resistance state corresponds to $\alpha = 180°$. A shift of the transverse resistance by an amount $R_{PHE}^{0°/180°}$ (cyan-dashed line) is observed at large magnetic fields where ***m*** and $I_{bias}$ are misaligned. The experimental spin Hall signal $2\Delta R_{ISHE}^{exp}$ is the same as the spin Hall signal obtained from the fit on the angle dependence to the transverse resistance $2\Delta R_{ISHE}^{fit}$.

**Appendix 2: The ordinary Hall effect in the CoFe/Pt nanostructure**

The OHE in Pt is yet another Hall effect that could compete with the ISHE in the CoFe/Pt nanostructure. The magnetization at the tip of the CoFe electrode can induce a magnetic stray field with a z-component penetrating the Pt electrode. This magnetic field, together with the current along the x-axis, can yield an OHE signal in the y-axis, the same one along which the ISHE is measured. The ordinary Hall resistance, which can contaminate the spin Hall signal, is $R_{xy}^{OHE} = R_H B_z/t_{Pt}$, with $R_H, B_z$ and $t_{Pt}$ being the material-dependent Hall coefficient, the z-component of the magnetic field in the Pt wire and the thickness of the Pt electrode, respectively. We combine electronic Hall measurements (to obtain $R_H$) with micromagnetic simulation (to extract the mean stray field component $<B_z^{stray}>$ created by the CoFe electrode) to show that the signal induced by the OHE is negligible in comparison to the ISHE.

To quantify $R_H$ in our Pt wire, an angular-dependent Hall measurement with a fixed magnetic field ($H$) of 50 kOe is performed in a Hall cross of Pt that is deposited at the same time as the CoFe/Pt nanostructure, as shown in Fig. 5(a). Figure 5(b) presents the resulting Hall resistance as a function of the out-of-plane rotation in the $xz$-plane (described by the angle $\beta$). We assume that the OHE in our Pt is linear in the out-of-plane magnetic field, as confirmed by the excellent fit (red line) of the measurement to $R_{xy}^{OHE}(\beta) = R_{xy}^{OHE}\cos(\beta + \beta_0) + b$ with $R_{xy}^{OHE} = (15.85 \pm 0.05)$ mΩ, $\beta_0 = -(0.5 \pm 0.2)°$ and $b = -(0.434 \pm 0.03)$ mΩ. The Hall coefficient is $R_H = R_{xy}^{OHE} t_{Pt}/B_z = (2.5 \pm 0.8) \times 10^{-11}$ m³/C as $t_{Pt} = 8$ nm and $B_z = 5$ T defined by fixed value of the external magnetic field $\mu_0 H$ ($\mu_0$ being the vacuum permeability).

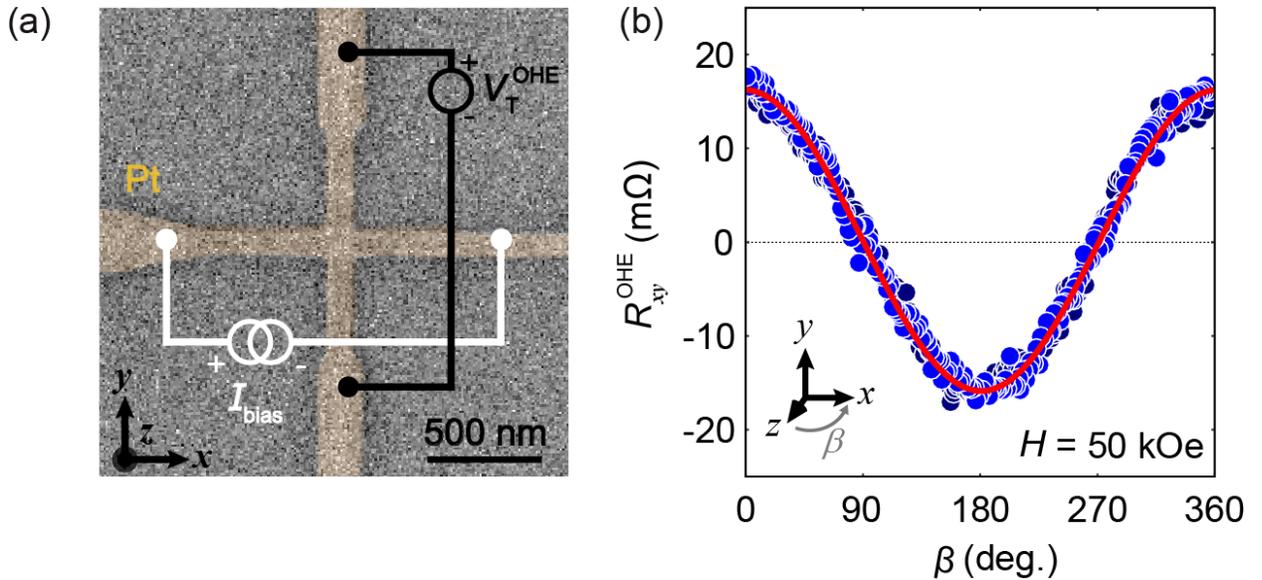

FIG. 5: (a) The ordinary Hall measurement configuration at the Pt Hall cross fabricated next to the local spin injection device. (b) The angular dependence of the Hall resistance ($R_{xy}^{OHE}$) as a function of the angle $\beta$ with a fixed magnetic field of 50 kOe. A fit to a $\cos(\beta)$ function is show as a red line.

The magnitude of the stray fields originating from the CoFe electrode into the Pt electrode are determined using MuMax3, which is an open-source software that allows for GPU-accelerated micromagnetic simulation.[40] Figure 6(b) shows the CoFe/Pt structure used for these simulation considering $t_{Pt} = 9$ nm, $t_{CoFe} = 15$ nm, $w_{Pt} = 215$ nm, $w_{CoFe} = 185$ nm, comparable to the dimension in the CoFe/Pt nanostructure studied in the main text. The wires are divided into a regular mesh of cuboid cells with dimensions of 3.125 nm x 3.125 nm x 3 nm. For this reason, the thickness of the Pt electrode in the simulation is chosen to be 1 nm thicker than the real electrode thickness. This, however, will not influence the result much since the strongest stray field will be induced in the area closest to FM electrode. A saturation magnetization of $M_{\text{sat}}=1$ MA/m (the value for non-annealed CoFe)[41] and an exchange constant $A_{\text{ex}}= 18$ pJ/m (the value for very thin Co)[42] have been used for the simulations. We consider that there is no anisotropy, since our CoFe is polycrystalline.

First, we simulate the three components ($x$, $y$, and $z$) of the magnetization in the CoFe electrode and its response to an external magnetic field sweep along the $x$-axis to imitate the ISHE measurement. Figure 6(a) shows the average magnetization in the area where the CoFe wire overlaps with the Pt electrode versus the magnitude of the external magnetic field, showing a coercive field of $H_c \sim 950$ Oe. The average magnetization in the $x$-direction is the strongest, as expected. However, around $H_c$, a component in the $z$-direction appears while the $y$-component stays rather constant in the full range of external magnetic field. Figure 6(b) presents a top view of the simulated structure, including snapshots of the magnetization at two different magnetic fields around $H_c$, which are displayed as blue dots in Figure 6(a).

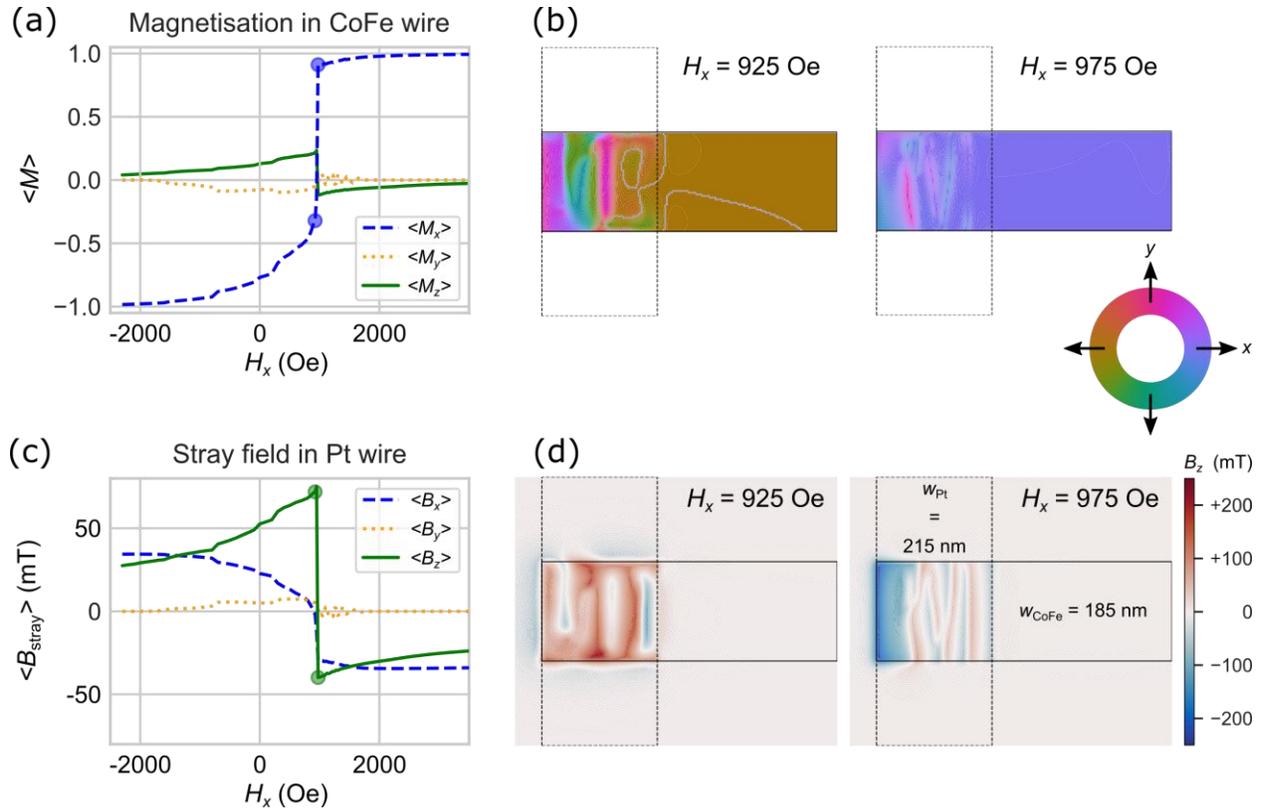

FIG. 6: (a) The simulated mean magnetization components of CoFe in the area where the CoFe wire overlaps with the Pt wire as a function of a field $H_x$ applied along the CoFe wire long axis. (b) Snapshots of the in-plane magnetization (see color-wheel) at a height of $z = 9$ nm, which corresponds to the vertical middle of the CoFe electrode ($t_{\text{CoFe}} =15$ nm, with the 3 nm cell height), at specific fields during the switching of the magnetization. The location of the snapshots in magnetic field are indicated by the blue

dots in panel (a). (c) Mean stray-field components generated in the Pt wire overlapping with the CoFe wire as a function of a field $H_x$. The maximum value of $<B_{\text{stray},z}>$ reaches ~75 mT and saturates to about 25 mT at high saturation fields. (d) Snapshots of the $z$-component of the stray field in the Pt wire at $z = 21$ nm (i.e. 6 nm above the CoFe wire) evaluated at the same fields as the CoFe magnetization in panel (b). The green dots in panel (c) indicate the magnetic field position of the snapshot.

Next, we simulate the stray field in the Pt electrode induced as a consequence of the magnetization in the CoFe wire. Figure 6(c) shows that the CoFe magnetization induces an average stray field in the Pt electrode with strong components in the $x$-direction and $z$-direction. The $z$-component is the one we are interested in, as it will induce a transverse resistance due to the OHE in the Pt which adds to the spin Hall signal. We find that the average stray field in the z-direction reaches a maximum $<B_{\text{stray},z}> \sim 75$ mT close to the coercive field $H_c \sim 950$ Oe. For higher applied fields, $<B_{\text{stray},z}>$ saturates with a value of ~25 mT. Figure 6(d) presents a top-view of the CoFe/Pt nanostructure with the $z$-component of the stray field at the intersection area of both wires.

The Hall resistance which will appear in the ISHE measurement, i.e. that will contaminate the spin Hall signal, can be estimated by combining the Hall coefficient ($R_H = 2.5 \times 10^{-11}$ m$^3$/C), the Pt wire thickness $t_{\text{Pt}}$ and the $z$-component of the magnetic stray field within the Pt wire. The maximum contribution of the OHE will be around the switching field, where $<B_{\text{stray},z}> \sim 95$ mT results in an OHE resistance of 0.3 mΩ, which is insignificant. The contamination of the transverse resistance should be considered at the field of saturation where the stray field is $<B_{\text{stray},z}> \sim 25$ mT, which yields an OHE resistance of 0.08 mΩ. This is about 0.3% of the total transverse resistance [(24±2) mΩ], meaning that contamination of the ISHE with the OHE is negligible.

**Appendix 3: Expected shape of transverse resistance loop in the presence of the PHE**

The PHE appears only in the transverse resistance when there is misalignment between the applied current $I_{\text{bias}}$ and the magnetization $\boldsymbol{m}$. In the ISHE measurement setup, $\boldsymbol{m}$ is aligned along the easy axis of the FM electrode at low magnetic fields, and along the external magnetic field $\boldsymbol{H}$ at high fields. Therefore, at low fields, $\boldsymbol{m}$ will be parallel to $I_{\text{bias}}$, whereas, at high fields, a misalignment between $\boldsymbol{m}$ and $I_{\text{bias}}$ can be introduced due to the positioning of the device on the sample during fabrication and/or the placement of the sample inside the measurement station. Figure (7) shows a sketch of $I_{\text{bias}}$, $\boldsymbol{m}$ and $\boldsymbol{H}$ for negative and positive misalignment angles ($\alpha_0$) in our ISHE measurement configuration.

The angle between $I_{\text{bias}}$ and $\boldsymbol{m}$ is given by $\varphi = \alpha + \alpha_0$, where $\alpha$ is the angle between the sample holder and $\boldsymbol{H}$ defined by the equipment. For the $R_T^{\text{ISHE}}$ versus $\boldsymbol{H}$ measurements, $\alpha$ is set to 180°. As shown in Eq. (2), the ISHE resistance and the PHE resistance are given by $R_{\text{ISHE}} = a_{\text{ISHE}} \cos(\varphi)$ and $R_{\text{PHE}} = a_{\text{PHE}} \sin(2\varphi)$, respectively. At low magnetic field (yellow area), $I_{\text{bias}}$ and $\boldsymbol{m}$ are aligned because $\boldsymbol{H}$ is not strong enough to overcome the shape anisotropy, which aligns $\boldsymbol{m}$ along the easy axis of the FM electrode. In this situation, $\varphi = 0$, i.e., $R_{\text{PHE}}$ is zero and, therefore, we only have a contribution of the ISHE resistance. However, when we increase $\boldsymbol{H}$ such that $\boldsymbol{m}$ aligns with $\boldsymbol{H}$ (green area), the misalignment angle is also induced between $I_{\text{bias}}$ and $\boldsymbol{m}$ ($\varphi = \alpha_0$) and $R_{\text{PHE}}$ becomes nonzero. The PHE resistance is negative or positive, depending on whether $\alpha_0$ is negative [Fig. 7(a)] or positive [Fig. 7(b)], respectively.

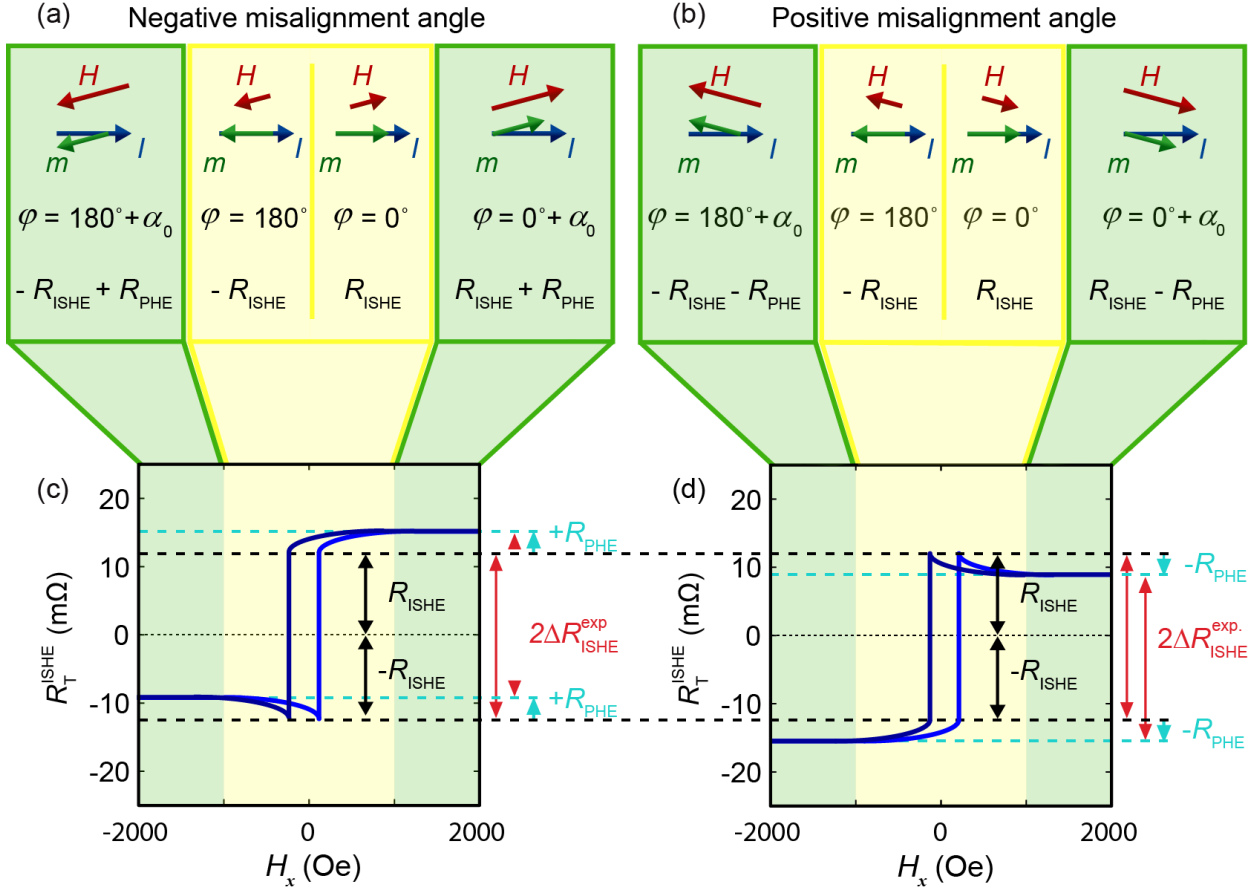

FIG. 7: Expected shape of the transverse resistance $R_T^{ISHE}(H)$ loop induced by the PHE contribution. (a) and (b) Sketch of the alignment of the current $I_{bias}$, the magnetization $m$ and the external magnetic field $H$ at high (green areas) and low magnetic field (yellow areas) for a negative and positive misalignment angle $\alpha_0$ between $I_{bias}$ and $m$. When $I_{bias}$ and $m$ are parallel, the transverse resistance is equal to the ISHE resistance, i.e. $R_{ISHE}$, if no other Hall effects are involved. If there is a misalignment between $I_{bias}$ and $H$, the misalignment is transferred to $m$ above a certain threshold value of $H$, therefore the PHE contribution $R_{PHE}$ appears at magnetic fields above this threshold value only. (c) and (d) $R_T^{ISHE}(H)$ loop as a function of the applied magnetic field in the ISHE configuration, for negative and positive misalignment angles, respectively. The effect of the PHE on $R_T^{ISHE}$ is two-fold. First, the PHE shifts the baseline signal down (negative misalignment angle) or up (positive misalignment angle). Secondly, the PHE induces specific shapes (dips and upturns) close to the switching field of the magnetization in the magnetic field dependence.

Figures 7(c) and 7(d) display a sketch of the transverse resistance $R_T^{ISHE}(H)$ as a function of the magnetic field. In the ideal case, without misalignment and vanishing PHE contribution, the switching of the resistance states would be sharp and flat (square loop). However, in the presence of a misalignment angle, the shape of the $R_T^{ISHE}$ loop is altered because of the dependence on the strength of $H$ as discussed before. Ideally, when sweeping $H$ from negative to positive and considering a negative misalignment angle [see Fig. 7(c)], the transverse resistance starts at a value equal to $-R_{ISHE} + R_{PHE}$, which slowly decreases to $-R_{ISHE}$ because $m$ rotates towards the easy axis of the FM electrode, that is parallel to $I_{bias}$, as $H$ moves to positive values. This creates the sharp dip in the $R_T^{ISHE}$ loop. By further increasing $H$, $m$ switches 180° but will be still parallel to $I_{bias}$ resulting in a transverse resistance of $R_{ISHE}$. Finally, at high magnetic field values, $m$ aligns again along **H** and the transverse resistance obtains an additional contribution $+R_{\text{PHE}}$. When sweeping from positive to negative magnetic field values, the $R_{\text{T}}^{\text{ISHE}}(H)$ curve has same shape but is shifted by the magnetic hysteresis. Figure 7(d) shows a similar behavior, although inverted because the positive misalignment angle induces a negative PHE resistance ($-R_{\text{PHE}}$). This specific shape is not precisely observed in our experimental measurement which is most probably related to the fact that the magnetization in the tip of the FM is not perfectly aligned along the easy axis of the FM at low magnetic field, as also observed in the micromagnetic simulations in Figs. 6(a) and 6(b). Nevertheless, both for negative as well as positive misalignment angles, the PHE resistance has the same magnitude and sign at high positive and negative magnetic fields Correspondingly, the spin Hall signal ($\Delta R_{\text{ISHE}}^{\text{exp}}$) can be accurately extracted from the difference between the transverse resistance at field values above saturation of the PHE.

Note that the sketch in Fig. 7 is made considering a material with a positive spin Hall angle, such as Pt used here. Materials with a negative spin Hall angle, such as Ta and W, will have a positive and negative $R_{\text{T}}^{\text{ISHE}}$ value at negative and positive magnetic field values, respectively, opposite to Pt. However, in the presence of a misalignment angle, the PHE characteristics in the $R_{\text{T}}^{\text{ISHE}}$ loop (the dips and upturns around the switching fields as presented here for Pt) will be equally valid in negative spin Hall materials.